\documentclass{article}  
\usepackage{bigsky2007}
\usepackage{graphics}
\usepackage{graphicx}
\usepackage[figuresright]{rotating}
\usepackage{epsfig}
\usepackage{subfigure}
\usepackage{url}
\usepackage{hyperref}
\newcommand{\pion}{$\pi^0$ }
\newcommand{\epair}{e$^+$e$^-$--pair }
\newcommand{\epairs}{e$^+$e$^-$--pairs }
\newcommand{\pt}{$p_T$ }

\newcommand{\AuAu}{$\mathrm{Au} + \mathrm{Au}$ }

\frompage{000} \topage{000}                                              
\def\rightmark{Measurement of photons via conversion pairs with PHENIX
at RHIC}
\def\leftmark{T. Dahms}
\title{Measurement of photons via conversion pairs in Au$\mathbf{+}$Au
  collisions at $\sqrt{s_{NN}}$~=~200~GeV with the PHENIX experiment
  at RHIC}
\authors{
  {T. Dahms$^1$ for the PHENIX collaboration
  }\\[2.812mm]
  {\normalsize
    \hspace*{-8pt}$^1$ Department of Physics and Astronomy, Stony Brook University,\\ 
    Stony Brook, NY 11794-3800, USA\\[0.2ex]
}}
  
\abstract{Thermal photons can provide information on the temperature
  of the new state of matter created at RHIC. In the \pt region of
  1--3~GeV/c thermal photons are expected to be the dominant direct
  photon source. Therefore, a possible excess compared to a pure decay
  photon signal due to a thermal photon contribution should be seen in
  the double ratio
  $(\gamma/\gamma(\pi^{0}))_\mathrm{Measured}/(\gamma/\gamma(\pi^{0}))_\mathrm{Simulated}$,
  if sufficient accuracy can be reached. We present a method to
  reconstruct direct photons by measuring \epairs from external photon
  conversions.}
  

\begin{document}
\maketitle
\setcounter{page}{1}
\section{Introduction}
\label{sec:introduction}
Direct photons are produced during all stages of heavy ion collisions
at the Relativistic Heavy Ion Collider (RHIC). Because they do not
interact strongly, they escape the medium unaffected by final state
interactions and provide a promising signature of the earliest and
hottest stage of the quark-gluon plasma (QGP)~\cite{Turbide:2003si}.

On a microscopic level, the main sources of direct photons from a QGP
are quark-gluon Compton scattering ($q g \rightarrow \gamma q$),
quark-anti\-quark annihilation ($q \bar{q} \rightarrow \gamma g$) and
brems\-strahlung involving thermalized
partons~\cite{Aurenche:1998nw}. Direct photons are also produced in
initial hard scattering processes which involve the same reactions but
among the incoming partons.

At RHIC energies thermal photons are predicted to be the dominant
source of direct photons in a \pt window between
1--3~GeV/c~\cite{Turbide:2003si}.

Direct photons have been measured with PHENIX in \AuAu collisions at
$\sqrt{s_{NN}} = 200$~GeV~\cite{Adler:2005ig}. The inclusive photon
spectra measured with the Electromagnetic Calorimeter (EMC) have been
compared to the expected background from hadronic sources, based on
the measured \pion and $\eta$ spectra and a cocktail of other hadronic
decays ($\eta^{\prime}$, $K_S^0$, $\omega$), assuming $m_T$ scaling.
\begin{figure}[t]
  \centering
  \includegraphics[width=0.85\textwidth]{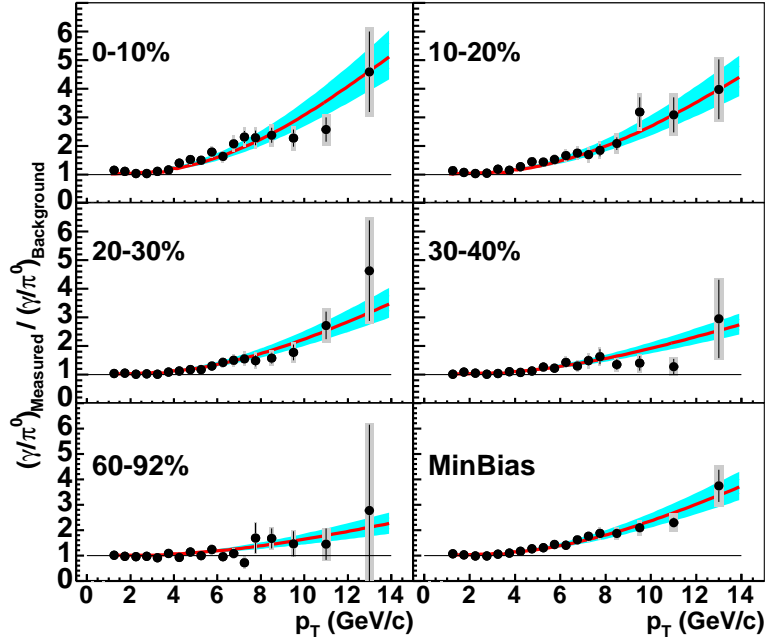}
  \caption{Double ratio of the measured invariant yield ratio,
    $\left(\gamma/\pi^0\right)_\mathrm{Measured}$, to the background
    decay ratio, $\left(\gamma/\pi^0\right)_\mathrm{Background}$, as a
    function of \pt for minimum bias and for five centralities of
    \AuAu collisions at $\sqrt{s_{NN}} = 200$~GeV.}
  \label{fig:doubleratio}
\end{figure}

Fig.~\ref{fig:doubleratio} shows the double ratio of the measured
invariant yield ratio to the background decay ratio as a function of
\pt for minimum bias and for five centrality classes. The measurement
of direct photons production at high \pt scales with the number of
binary collisions in agreement with NLO perturbative QCD predictions
and therefore confirms medium effects as the origin of jet
suppression. In the low \pt region, where a thermal signature is
expected, a significant measurement remains limited by systematic
uncertainties due to the energy resolution and the photon
identification with the EMC.

In order to overcome such limitations, dielectron pairs offer some
advantages because of the superior momentum resolution of charged
particles at low momenta and excellent identification of conversion
photons; while other methods~\cite{Bathe:2005nz} try to use low mass
dielectron pairs from internal conversions, the method presented here
uses real photon conversions in the beam pipe.

\section{Thermal photon analysis}
\label{sec:analysis}
The excellent capabilities of the PHENIX detector to measure electrons
suggest to circumvent the limitations of the conventional direct
photon measurement~\cite{Adler:2005ig} at low photon energies by
measuring photons via their conversion into $e^+e^-$--pairs. The
momentum resolution ($\sim 1\,\%$) of the charged tracking devices
proves superior to the energy resolution of the EMC ($\sim 10\,\%$) in
the \pt region of interest (1--3~GeV/c).

To identify \epairs from photon conversions, a single electron
identification cut is applied, which require signals from at least two
phototubes in the Ring Imaging Cherenkov Detector (RICH) matching to a
reconstructed charged track in the Drift Chamber (DC). No further
electron identification cuts were applied since the pair cuts (see
Sect.~\ref{sec:conversions}) to separate conversion photons from other
\epairs are more efficient and powerful enough to provide a very clean
photon conversion sample.

The extracted photon conversions are tagged with photons reconstructed
in the EMC to determine the contribution from $\pi^0 \rightarrow
\gamma \gamma$ decays (see Sect.~\ref{sec:tagging}).

All yields are measured as a function of $p_T$ of the
e$^+$e$^-$--pair, which makes a direct comparison of the inclusive
photon yield, $N_{\gamma}^\mathrm{incl}$, and the tagged photon yield,
$N_{\gamma}^{\pi^0 \mathrm{tag}}$, possible:
\begin{eqnarray}\label{eq:inclusive}
N_{\gamma}^\mathrm{incl}\left(p_T\right) &=&
\epsilon_{e^+e^-}~a_{e^+e^-}~\gamma^\mathrm{incl}\left(p_T\right) \\
\label{eq:piontag}
N_{\gamma}^{\pi^0 \mathrm{tag}}\left(p_T\right) &=&
\epsilon_{e^+e^-}~a_{e^+e^-}~
\epsilon_{\gamma}\left(p_T\right)~f~\gamma^{\pi^0}\left(p_T\right)
\end{eqnarray}
The measured yield of inclusive photons depends on the reconstruction
efficiency $\epsilon_{e^+e^-}$ and the PHENIX acceptance $a_{e^+e^-}$
of the conversion \epair. The tagged photon yield depends in addition
on the efficiency to reconstruct the second photon in the EMC
$\epsilon_{\gamma}(p_T)$ and on the conditional probability $f$ to
find it in the EMC acceptance, given that the \epair has been
reconstructed already. Here, $\epsilon_{\gamma}(p_T)$ is weighted with
the \pt distribution of the $e^+e^-$--pair. In the ratio
$N_{\gamma}^{incl}/N_{\gamma}^{\pi^0 tag}$ the \epair reconstruction
efficiency and acceptance correction factor cancel.

A ratio of the hadronic decay photon yield,
$N_{\gamma}^{\mathrm{hadr}}$, and the tagged photon yield from $\pi^0$
decays, $N_{\gamma}^{\pi^0 \mathrm{tag}}$, is calculated with
simulations.
\begin{eqnarray}\label{eq:pions}
N_{\gamma}^{\pi^0 \mathrm{tag}}\left(p_T\right) &=& f~N_{\gamma}^{\pi^0}\left(p_T\right)
\end{eqnarray}

The comparison of the ratio in data and in simulations in a double
ratio leads to an expression that is equivalent to the ratio of
inclusive and decay photons as shown in Eq.~(\ref{eq:doubleratio}).
\begin{eqnarray}\label{eq:doubleratio}
  \frac{\gamma^\mathrm{incl}\left(p_T\right)}{\gamma^\mathrm{hadr}\left(p_T\right)} =
  \frac{\epsilon_{\gamma}\left(p_T\right) \cdot
    \left(\frac{N_{\gamma}^\mathrm{incl}\left(p_T\right)}{N_{\gamma}^{\pi^0 \mathrm{tag}}\left(p_T\right)}\right)_\mathrm{Data}}
       {\left(\frac{N_{\gamma}^\mathrm{hadr}\left(p_T\right)}{f~N_{\gamma}^{\pi^0}\left(p_T\right)}\right)_\mathrm{Sim}}
\end{eqnarray}
The only remaining factors are the reconstruction efficiency of the
photon in the EMC, $\epsilon_{\gamma}(p_T)$, and the conditional
acceptance $f$ in the simulation part of the double ratio, which have
both been determined with Monte Carlo simulations (see
Sect.~\ref{sec:simulations}).

\subsection{Photon Conversions}
\label{sec:conversions}
Since the PHENIX tracking algorithm assumes the track to originate
from the collision vertex, off-vertex conversion pairs are
reconstructed with an artificial opening angle which leads to an
invariant mass that is proportional to the radius at which the
conversion occurs.

\begin{figure}[t]
  \centering
  \includegraphics[width=0.85\textwidth]{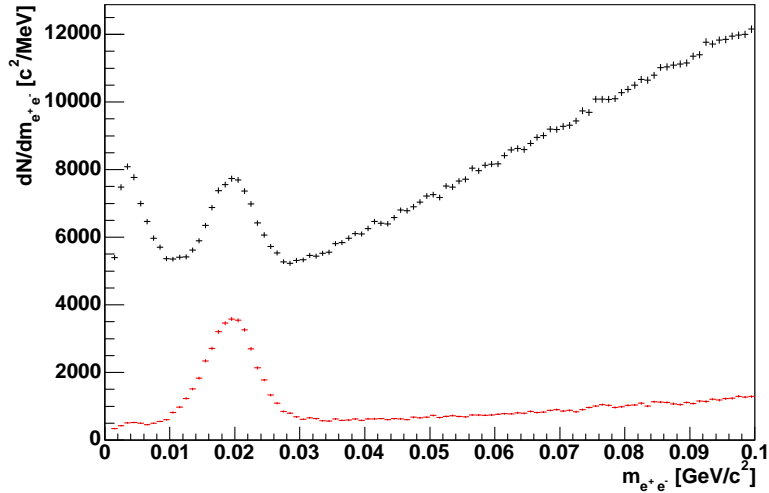}
  \caption{Invariant mass of \epairs before (black) and after (red)
    applying cuts on the orientation of the \epair in the magnetic
    field.}
  \label{fig:allpairs}
\end{figure}

Therefore, photon conversions that occur in the beam pipe material
(Be, $0.3\,\%$ radiation length) at a radius of 4~cm are reconstructed
with an invariant mass of $\sim
20~\mathrm{MeV/c^2}$. Fig.~\ref{fig:allpairs} shows an invariant mass
spectrum of \epairs in the range 0--0.1~GeV/c$^2$. The peak from
photon conversions in the beam pipe at 20~MeV/c$^2$ can be clearly
separated from Dalitz decays $\pi^0 \rightarrow \gamma e^+ e^-$, which
dominate the spectrum below 10~MeV/c$^2$, and combinatorial background
pairs, whose contribution increases toward higher invariant masses.

The photon conversion pairs, which have no intrinsic opening angle,
can be distinguished from Dalitz decays and purely combinatorial pairs
by cutting on the orientation of the \epair in the magnetic field.

Fig.~\ref{fig:allpairs} shows the invariant mass spectra of \epairs
before (black) and after (red) applying these pair cuts. The yield
from integrating the mass region $ < 35~\mathrm{MeV/c}^2$ of the
conversion peak is corrected for the remaining $p_T$ dependent
contamination of $\sim 15.0 \pm 2.0~\mathrm{(syst)}\,\%$ due to
combinatorial \epairs which has been determined with mixed events.

\subsection{Tagging of Decay Photons}
\label{sec:tagging}
\begin{figure}[t]
  \centering
  \includegraphics[width=0.85\textwidth]{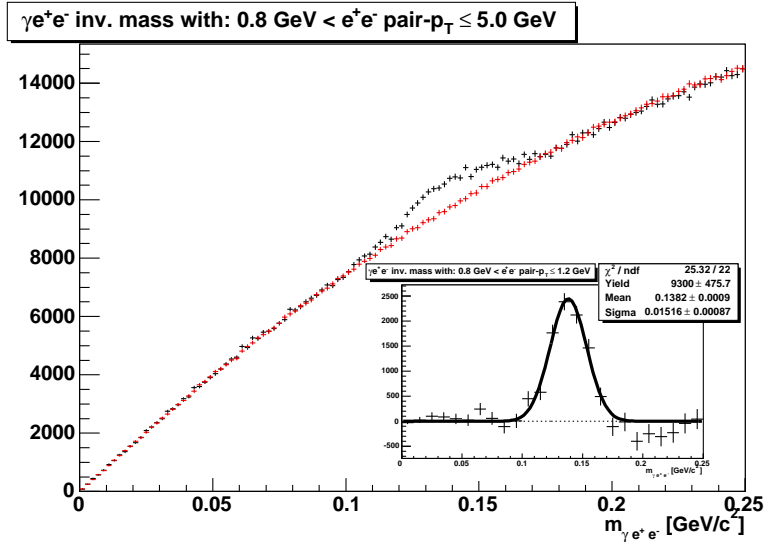}
  \caption{Invariant mass of $\gamma e^+ e^-$--triplets in same events
    (black) and normalized mixed events (red) for \epairs with
    $0.8~<~p_T~\leq~5.0$~GeV/c. The insert shows the invariant mass of
    $\gamma e^+ e^-$--triplets after background subtraction for
    \epairs with $0.8 < p_T \leq 1.2$~GeV/c. A fit with a Gaussian is
    drawn and the resulting parameters shown in the box in the upper
    right of the graph.}
  \label{fig:invmasstriplets}
\end{figure}

To reveal which of these conversion photons come from $\pi^0
\rightarrow \gamma \gamma$ decays, the \epairs in the conversion peak
are combined with photons which have been measured in the EMC, under
loose cuts based on the time of flight and the shower profile for
photons with a minimum \pt of $0.3~\mathrm{GeV/c}$, and their
invariant mass is calculated (see Fig.~\ref{fig:invmasstriplets}).

The reconstruction efficiency $\epsilon_{\gamma}(p_T)$ of the loose
photon has been estimated with a full GEANT simulation which embed
simulated photons into real EMC data, therefore providing a combined
information on the photon identification efficiency and occupancy
effects. The overall efficiency is determined to be $82 \pm 1\,\%$
independent of \pt beyond the minimum \pt cut off.

Conversion photons that are identified as decay products of $\pi^0$
can be tagged as $N_{\gamma}^{\pi^0 \mathrm{tag}}$. This signal has a
large combinatorial background due to the high photon multiplicity in
\AuAu collisions.

The combinatorial background is reproduced with an event mixing
method, which creates uncorrelated pairs of photons and \epairs from
different events. The mixed event spectrum is normalized to the same
event spectrum well outside the $\pi^0$ mass region (0--100~MeV/c$^2$,
170--250~MeV/c$^2$) and subtracted.

The statistical error on the normalization factor is on the order of
$0.2\,\%$ and depends only on the statistics in the same event
spectrum in the normalization region. As an example, the resulting
$\pi^0$ signal for \epairs with $0.8 < p_T \leq 1.2$~GeV/c is shown as
insert in Fig.~\ref{fig:invmasstriplets}.

Mean and $\sigma$ are determined by a fit of the background subtracted
data with a Gaussian. The data are also fitted to the sum of a second
order polynomial and a Gaussian, to take into account the possibility
that the shape is not completely described by the mixed event
spectrum. The difference in the resulting mean and $\sigma$ is
negligible. The mean and $\sigma$ obtained by the fit are then used to
integrate the data in a region $\pm 1.5~\sigma$ around the mean,
chosen to optimize the signal to background ratio.

The statistical error on the extracted $\pi^0$ signal is given by:
\begin{eqnarray}
\sigma_S^2 &=& \sum_i FG(i) + \alpha \sum_i BG^{\prime}(i) +
  \left(\frac{\sigma_{\alpha}}{\alpha} \sum_i
  BG^{\prime}(i)\right)^2\label{eq:staterror}
\end{eqnarray}
With $FG(i)$ and $BG^{\prime}(i)$ being the yields in bin $i$ of
invariant mass spectrum in same events and normalized mixed events,
respectively, the summations are performed over the integration
region. It is important to note that the last term in
Eq.~(\ref{eq:staterror}), is the square of the sum over the normalized
background, and therefore, depends on the integration region and is
not bin independent. Different integration regions have been
used. Variations in the resulting yield have been used to set a
systematic uncertainty on the yield extraction of $2.5\,\%$
independent of $p_T$.

The loss of $N_{\gamma}^{\pi^0 \mathrm{tag}}$ due to the external
conversion of the second photon is corrected by a factor $1-p_{conv} =
94 \pm 2\,\%$. In this factor $p_{conv}$ is the conversion probability
due to the material budget between the vertex and the Pad Chamber 3
(PC3) in front of the EMC.

\subsection{Simulations}
\label{sec:simulations}
The contribution of hadronic decays has been determined with a fast
Monte Carlo simulation of $\pi^0$ and $\eta$ Dalitz decays. A
parameterization of the $\pi^0$ spectrum measured by
PHENIX~\cite{Adler:2003qi} has been used as input. The $\eta$
distribution has been generated assuming $m_T$ scaling ($p_T
\rightarrow \sqrt{p_T^2 + m_{\eta}^2 - m_{\pi^0}^2}$) of the $\pi^0$
spectral shape and a normalization at high \pt to $\eta / \pi^0 = 0.45
\pm 0.04$, according to PHENIX
data~\cite{Adler:2006hu,Adler:2004ta}. The relative error of $9\,\%$
on the $\eta/\pi^0$ ratio is reduced by the branching ratio of the two
photon decay and results in a $3\,\%$ error in the ratio
$N_{\gamma}^{\mathrm{hadr}}/N_{\gamma}^{\pi^0}$.

The contamination due to neutral Kaons which decay before the beam
pipe has been found negligible ($\sim 1\,\%$) and has been folded into
the systematic error on the simulations.

The conditional probability $f$ that the photon is reconstructed in
the EMC once the \epair is reconstructed already was calculated with
a fast Monte Carlo simulation of $\pi^0 \rightarrow \gamma e^+
e^-$. The use of Dalitz decays is justified by the fact that the \pt
spectra of photons from $\pi^0 \rightarrow \gamma \gamma$ are
essentially identical to the \epair \pt spectrum from $\pi^0
\rightarrow \gamma e^+ e^-$.

After the \epair has been filtered in the detector acceptance, the
conditional acceptance $f$ for the second photon has been calculated
taking dead areas of the detector into account. Uncertainties in
calculating $f$ are found to be $5\,\%$, which is the largest source
of systematic errors. The limited energy resolution of the EMC of
$\sigma_E/E = 5\,\%~\oplus~9\,\%/\sqrt{E}$ introduced an additional
systematic error of $\sim 1\,\%$ due to the \pt cut at
$0.3~\mathrm{GeV/c}$.

\section{Conclusions} 
\label{sec:conclusions}
\begin{figure}[t]
  \centering
  \includegraphics[width=0.85\textwidth]{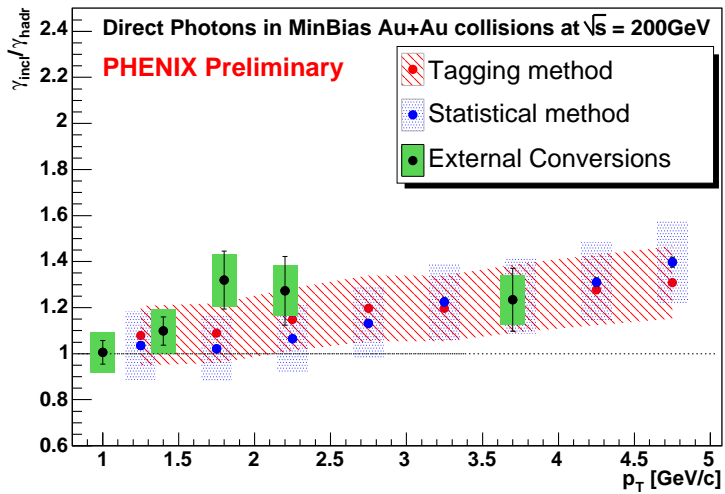}
  \caption{Ratio of inclusive photons to decay photons,
  $\gamma^\mathrm{incl}\left(p_T\right) /
  \gamma^\mathrm{hadr}\left(p_T\right)$, in Minimum Bias \AuAu
  collisions at $\sqrt{s_{NN}}$~=~200~GeV, for three different methods
  methods.}
  \label{fig:results}
\end{figure}

Fig.~\ref{fig:results} shows a preliminary result for the double ratio
$\gamma^\mathrm{incl}\left(p_T\right) /
\gamma^\mathrm{hadr}\left(p_T\right)$ as in Eq.~\ref{eq:doubleratio}
for Minimum Bias \AuAu collisions at $\sqrt{s_{NN}}$~=~200~GeV. The
main sources of systematic errors arise from the uncertainties in the
description of the detector active areas, in the peak extraction and
in the assumptions of the \pion shape and give a final systematic
error on the double ratio of $\sim 7\,\%$.

The result of the presented analysis is compared to two other direct
photon measurements in PHENIX. The first one~\cite{Gong:2007hr} is
based on the same tagging method, but instead of photons coming from
conversions in the beam pipe, the clean photon sample is determined by
selecting EMC clusters with very strict photon identification
cuts. The second one is the ratio obtained from the statistical
subtraction of measured \pion \pt spectra from the measured inclusive
photons~\cite{Isobe:2007ku}.

All three analyses indicate an excess due to direct photons above the
decay photon spectrum in the \pt region 1--3~GeV/c. Despite of the
statistical limitations due to the low conversion probability of
$0.2\,\%$ this analysis offers a smaller systematic error with respect
to the conventional methods, and therefore, the intriguing ability to
extend the extraction of a significant direct photon yield to low
$p_T$.

\vfill\eject
\end{document}